\documentclass[twocolumn,showpacs,preprintnumbers,amsmath,amssymb]{revtex4}

\usepackage{graphicx}
\usepackage{dcolumn}
\usepackage{bm}

\begin{document}


\title{Quantum statistical model for superconducting phase in graphene and nanotubes}

\author{Shigeji Fujita}
\author{Rohit Singh}%
\affiliation{%
Department of Physics, University at Buffalo, SUNY, Buffalo, NY 14260, USA
}%

\author{Salvador Godoy}
\affiliation{
Departamento de F\'{\i}sica, Facultad de Ciencias,
Universidad Nacional Aut\'onoma de M\'exico,
M\'exico 04510, D.F., M\'exico.
}%

\author{Kei Ito}
\affiliation{
Research Division, The National Center for University Entrance Examinations,\\
2-19-23 Komaba, Meguro, Tokyo 153-8501, Japan
}

\date{\today}

\begin{abstract}
A quantum statistical theory is presented, supporting a superconducting state of an ultrahigh
critical temperature (1275 K) in the multiwalled nanotubes reported by Zhao and Beeli [Phys.\@ Rev.\@ B \textbf{77}, 245433 (2008)].
\end{abstract}

\maketitle

\section{Introduction}

Zhao and his group [1] found an experimental evidence of superconductivity in multi-walled nanotubes (MWNT) with the critical temperature $T_{c}=1275$ K after examining the paramagnetic Meissner effect [2].
This is a new record of high $T_{c}$, about eight times higher than $T_{c}=164$ K found in mercury-based cuprates [3].
Fujita and Godoy developed a quantum statistical theory of superconductivity [4], extending the Bardeen--Cooper--Shrieffer (BCS) theory [5] by incorporating lattice and band structures of electrons and phonons and considering moving Cooper pairs.
The Cooper pairs [6], called the pairons, are formed by the phonon-exchange attraction.
These pairons move with linear dispersion relations, see Eq.~(2) [4].
The centers of mass (CM) of pairons move as bosons [4].
These pairons undergo a Bose--Einstein condensation (BEC) in two dimensions (2D).
We use the quantum statistical theory to support a superconductivity in MWNT at high temperatures ($\sim 1275$).

Zhang \textit{et al}.\@ [7] showed that the graphene is supercunducting at 1.7 K.
We discuss this in Section II.
The graphene sheet can be wrapped into a carbon nanotube.
The ``holes" (and not the ``electrons") can go through inside the positively charged graphene wall.
Due to this extra channel the nanotube's majority carriers are ``holes" although the graphene's majority carriers are ``electrons", see Section III.
In MWNT the ``holes" are generated more abundantly between the graphene walls.
We discuss the superconductivity at ultrahigh critical temperatures in Section IV.
Summary and discussion are given in Section V.

\section{Graphene}

In graphene carbon ions (C$^{+}$) occupy the two-dimensional (2D) honeycomb crystal lattice.
We assume that the ``electron" (``hole") wave packet has a negative (positive) charge $-$ ($+$) $e$ and a size extending over the unit C-hexagon.
The positively charged ``hole" tends to stay away from positive ions C$^{+}$ and hence its charge is concentrated at the center of the hexagon.
The negatively charged ``electron" tends to stay close to the C$^{+}$ ions and its charge is concentrated near the C$^{+}$ hexagon.
Because of the different internal charge distributions, the ``electron" and ``hole" should have different effective masses $m_{e}$ and $m_{h}$.
For the description of the electron motion in terms of the mass tensor, it is convenient to introduce Cartesian coordinates, which may not necessarily match with the crystal's natural (triangular) axes.
The ``electron" may move easily with a smaller effective mass in the direction [110] than perpendicular to it as we see presently.
Here we use the conventional Miller indices for the hexagonal lattice with omission of the c-axis index.
We may choose the unit cell as shown in Fig.~1.
\begin{figure}[tb]
  \begin{center}
    \includegraphics*[width=5cm,clip]{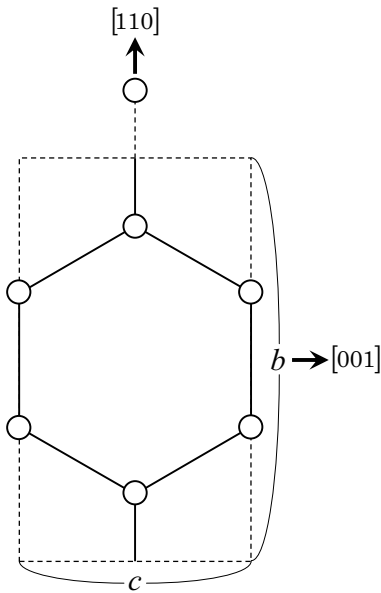}
    \caption{{\footnotesize
    A unit cell for graphene.
    }}
    \label{fig:1}
  \end{center}
\end{figure}
The choice is not unique.
But the size of the rectangle with side-length pair ($b$, $c$) for any unit cell is the same.
Then, the Brillouin zone in the $k$-space is unique: a rectangle with side length ($2\pi/b$, $2\pi/c$).
The ``electron" (wave packet) may move up or down to the neighboring hexagon sites passing over one C$^{+}$.
The positively charged C$^{+}$ acts as a welcome potential valley for the negatively charged ``electron".
The same C$^{+}$ acts as a hindering potential hill for the positively charged ``hole".
The ``hole" can move easily over on a series of vacant sites, each surrounded by six C$^{+}$, without meeting the hindering potential hills.
Thus, the easy channel direction for the ``electrons" and ``hole" are [110] and [001], respectively.

Let us consider the system (graphene) at 0 K.
If we put an electron in the crystal, then the electron should occupy the center O of the Brillouin zone, where the lowest energy lies.
Additional electrons occupy points neighboring O in consideration of Pauli's exclusion principle.
The electron distribution is lattice-periodic over the entire crystal in accordance with the Bloch theorem.
The graphene is quadrivalent metal.
The first few low-lying bands in the $k$-space are completely filled.
The uppermost partially filled bands are important for the transport properties discussion.
We consider such a band.
The Fermi surface which defines the boundary between the filled and unfilled $k$-spaces (areas) is not a circle since the $x$-$y$ symmetry is broken.
The effective mass $m_{e}$ is less in the direction [110] than perpendicular to it.
If the electron number is raised by the gate voltage, then the Fermi surface more quickly grows in the easy-axis (y-) direction than in the $x$-direction.
The Fermi surface must approach the Brillouin boundary at right angles because of the inversion symmetry possessed by the honeycomb lattice.
Hence, the Fermi surface must touch the Brillouin boundary at a certain voltage and a ``neck" Fermi surface must be developed.

The same easy channels in which the ``electron" runs with a small mass, may be assumed for other hexagonal directions, [011] and [101].
Hence, the system does not show anisotropy.

Experiments [7] indicate that (a) both ``electrons" and ``hole" can be excited in graphene, (b) at zero gate voltage the ``electrons" are dominant, (c) the resistivity $\rho$ exhibits a sharp maximum at the ``electrons" density $n_{e} \sim 2 \times 10^{11}$ cm$^{-2}$, and (d) the mobility $\mu$ proportional to the conductivity $\sigma$ shoot up at the ``hole" density $n_{h} \sim 3 \times 10^{11}$ cm$^{-2}$.
The feature (b) should arise from the existence of the welcoming ion C$^{+}$ for the ``electrons" (and not for the holes).
The feature (c) is due to the fact that the conductivity,
\begin{equation}
\sigma \equiv \rho^{-1} = e^{2} n_{e} \tau / m^{*},
\end{equation}
where $\tau$ is the relaxation time, must decrease since the effective mass $m^{*}$ shoots up to $\infty$ in the small-``neck" limit.
The feature (d) means that the graphene is superconducting at 1.7 K.
In our quantum statistical theory the pairons are generated through the phonon exchange attraction.
Since the phonon carries no charge, the system before and after the phonon exchange has the same charge state, requiring that positively and negatively charge pairons are created in equal numbers simultaneously.
The graphene's majority carriers are ``electrons" at zero gate voltage.
The mobility-maximum occurs on the ``hole" gate voltage side where the ``hole" density is high near the ``neck" Fermi surface.
The ``necks" can occur for both ``electrons" and ``holes" sides.
There is a  $\rho$-maximum for the ``electron" side and a $\mu$-maximum for the hole side at 1.7 K.
Why is the asymmetry?
The $\mu$-maxium means that the phonon exchange is in action.
There are not many ``holes" and hence the critical temperature $T_{c}$ is very low on the electron side.
If the system is observed at lower temperatures, it should then show superconductivity states on both sides. 

\section{Nanotube}

Let us now consider a long nanotube wrapped with the graphene sheet.
The charge may be transported by the channeling ``electrons" and ``holes" in the graphene wall.
But the ``holes" present inside the nanotube can also contribute to the charge transport.
The carbon ions in the wall are positively charged.
Hence, the positively charged ``hole" can go through inside the tube.
In contrast, the negatively charged ``electrons" are attracted by the carbon wall and cannot go straight through the tube.
Because the extra-channel inside the carbon nanotube,``holes" are the dominant charge carriers in the nanotube.
We predict the superconductivity states for the nanotube with $T_{c}$ greater than 1.7 K.

\section{Multiwalled nanotube}

A graphene sheat can be rolled like a wallpaper to produce a multiwalled nanotube (MWNT).
The tube diameter may reach 50 \AA.
Recently, Zhao and his group [3] found an evidence of superconductivity of a ultrahigh critical temperature (1275 K).

The pairons move in two dimensions (2D) with a linear dispersion relation
\begin{equation}
w_{p}^{(j)} = w_{0} + (2/\pi)v_{n}^{(j)}p,
\end{equation}
where $w_{0}$ is the negative ground state energy, $p$ the pairon momentum magnitude, and $v_{F}^{(j)}$ the Fermi velocity of type $j$, $j=1$ (``electron"), $j=2$ (``hole").
This relation is obtained, starting with a BCS-like Hamiltonian, setting up and solving an energy-eigenvalue problem for the moving pairon, see reference [4].
These pairons move similar to massless particles with the common speed ($2/\pi$) $v_{F}^{(j)}$.

In graphene the ``electron" pairons, having the greater speed, dominate the transport and the BEC.
The critical (superconducting) temperature $T_{c}$ is given by
\begin{equation}
k_{B}T_{c}=1.24 \, \hbar v_{F}^{(1)}n_{0}^{1/2},
\end{equation}
where $n_{0}$ is the pairon density.
Briefly the BEC occurs when the chemical potential vanishes at a finite $T$.
The critical temperature $T_{c}$ can be determined from
\begin{equation}
n_{0} = (2\pi\hbar)^{-2} \int d^{2}p \; [e^{\beta_{c} e}]^{-1}, \qquad \beta_{c} \equiv (k_{B}T_{c})^{-1}.
\end{equation}
After expanding the integrand in powers of $e^{-\beta_{c} e}$ and using $\epsilon=cp$, we obtain $n=1.654 \, (2\pi)^{-1} (k_{B}T_{c}/\hbar c)^{2}$, yielding formula (3) with $c=(2/\pi)v_{F}$.
Experiments [8] indicate that the linear coefficient $c$ equals $1.02 \times 10^{6}$ m/s, implying that the Fermi velocity $v_{F}$ is approximately equal to $1.57 \times 10^{6}$ ms$^{-1}$.
Using Eq.~(3), we obtain the critical temperature $T_{c}=1275$ K, for $n_{0}=7.33 \times 10^{11}$ cm$^{-2}$.
The measurments of the satulation magnetization of MWNT bundle [1] yield the 3D electron density $2.12 \times 10^{19}$ cm$^{-3}$, which give a 2D density $(2.12)^{2/3} \times 10^{12}$ cm$^{-2}$.
The pairon density must be smaller than the electron density.
Hence, the quoted $n_{0}=7.33 \times 10^{11}$ cm$^{-2}$ is reasonable

\section{Summary and Discussion}

We have shown, based on the quantum statistical theory, that a superconducting state exists in MWNT with a ultrahigh critical temperature ($\sim 1275$ K).
The linear dispersion relation in Eq.~(2) may be probed by the angle-resolved photo-emission spectroscopy (ARPES).
Lanzara \textit{et al}.\@ [9] applied ARPES to demonstrate the linear dispersion relation in the quantum Hall effect in graphene, which supports our theoretical formula (2).

\begin{acknowledgments}
The authors wish to thank Professors Guo-meng Zhao, Manuel de Llano and Murthy Ganapathy for enlightening discussions.
\end{acknowledgments}


\begin{thebibliography}{99}
\bibitem{1}
G-M. Zhao and P. Beeli, Phys.\@ Rev.\@ B \textbf{77}, 245433 (2008).
\bibitem{2}
B. I. Spivak and S. A. Kivelson, Phys.\@ Rev.\@ B \textbf{43}, 3740 (1991).
\bibitem{3}
Y. Maeno \textit{et al}., Nature \textbf{372}, 532 (1994).
\bibitem{4}
S. Fujita and S. Godoy, \textit{Quantum Statistical Theory of Superconductivity} (Plenum, New York, 1996), pp.\@ 167, 181-183, 184-185.
\bibitem{5}
J. Bardeen, L. N. Cooper, and J. R. Schrieffer, Phys.\@ Rev.\@ \textbf{108}, 1175 (1957).
\bibitem{6}
L. N. Cooper, Phys.\@ Rev.\@ \textbf{104}, 1189 (1956).
\bibitem{7}
Y. Zhang \textit{et al}., Nature \textbf{438}, 201 (2005).
\bibitem{8}
M. Orlita \textit{et al}., J. Phys:\@ Condens.\@ Matter \textbf{20}, 454223 (2008).
\bibitem{9}
S. Y. Zhou \textit{et al}., Nature Physics \textbf{2}, 595 (2006).
\end{thebibliography}
\end{document}